\documentstyle[12pt, aps]{revtex}

\begin{document}
\title{Multiple Perron-Frobenius operators}
\author{Yu. Dabaghian}
\address
{Department of Physics, Wesleyan University,
Middletown, CT 06459-0155
}
\address{ydabaghian@wesleyan.edu}
\date{May 1, 2000}

\maketitle

\begin{abstract}
A cycle expansion technique for discrete sums of several PF operators,
similar to the one used in standard classical dynamical zeta-function
formalism is constructed. It is shown that the corresponding expansion
coefficients show an interesting universal behavior, which illustrates the
details of the interference between the particlar mappings entering the sum.
\end{abstract}

\section{General introduction}

Using the formalism of dynamical zeta functions one can compute global
averages of the observables, associated with a fully chaotic dynamical
system 
\begin{equation}
\dot{x}_{i}=F_{i}(x).
\label{z1}
\end{equation}
In order to find the average for an observable $\Phi(x) $, one
defines a Perron-Frobenius operator, 
\begin{equation}
L_{\Phi}(x,y) =\delta \left(y_{\mu}-f_{\mu}(x)\right) 
e^{\beta \Phi (x)},  
\label{z2}
\end{equation}
where $f_{\mu}(x) $ is a flow associated with the system (\ref
{z1}), and $\beta $ is a parameter. The time average of $\Phi(x)$ can 
be found as 
\begin{equation}
\left\langle \Phi \right\rangle =\frac{\partial}{\partial \beta}\ln
z_{\min}(\beta) \mid _{\beta =0}, 
\end{equation}
where $z_{\min}(\beta) $ is the smallest root of the
eigenvalue equation 
\begin{equation}
\det \left(1-zL\right) =0.  
\label{z3}
\end{equation}

For fully chaotic (axiom A) systems the determinant (\ref{z3}) is an entire
analytic function of $z$, which can be written \cite[2, 3]{a1} in an
infinite product form as 
\begin{equation}
\det \left(1-zL\right) =\prod_{m}\prod_{p}\left(1-\frac{z^{n_{p}}
e^{\beta\Phi (x) n_{p}}}{\left| \Lambda _{p}\right| \Lambda _{p}^{m}}
\right) \equiv \prod_{m}\prod_{p}\left(1-z^{n_{p}}A_{p}^{n_{p}}t_{p}\right).
\end{equation}
Here $t_{p}\equiv \frac{1}{\left| \Lambda _{p}\right| \Lambda _{p}^{m}}$ is
the weight of the prime periodic orbit of the system (\ref{z1}) indexed by 
$p$, $\Lambda _{p}$ is it's Lyapunov exponent, $A_{p}$ is a short notation
for the amplitude, and $m$ enumerates repetitions of periodic orbits. The
product over the prime periodic orbits only, 
\begin{equation}
\zeta_{m}^{-1}(z) =\prod_{p}\left(1-z^{n_{p}}A^{n_{p}}t_{p}\right) ,
\label{z4}
\end{equation}
is known \cite{a2} as the dynamical zeta function associated with the system
(\ref{z1}), which is a meromorphic function in a certain domain of the
complex variable $z$ \cite[2, 3]{a1}.

An important feature of the theory based on the dynamical zeta function
considerations, is the possibility to determine the coefficients of the
power expansion of this function, 
\begin{equation}
\zeta _{m}^{-1}=1-t_{f}z-c_{2}z^{2}-c_{3}z^{3}-...,
\end{equation}
using very effective and insightful ``cycle expansion'' technique \cite{a4}.
As an example, in the case of a system whose orbits can described by a
simple binary code, one can write the following cycle expansion of the zeta
function $\zeta _{m}^{-1}$: 
\begin{eqnarray*}
\zeta _{m}^{-1} &=&\left(1-zt_{0}\right) \left(1-zt_{1}\right) 
\left(1-z^{2}t_{01}\right) \left(1-z^{3}t_{001}\right) 
\left(1-z^{3}t_{011}\right) ... \cr
&=&\allowbreak 1-\left(t_{1}+t_{0}\right) z-\left(t_{01}-t_{0}t_{1}\right)
z^{2}-\left(\left(t_{001}-t_{0}t_{01}\right) 
+\left(t_{011}-t_{1}t_{01}\right) \right) z^{3}+....
\end{eqnarray*}
It is easy to observe that in the ``curvature coefficients'', which are in
this case 
\begin{eqnarray}
c_{2} &=&t_{01}-t_{0}t_{1}, \cr
c_{3} &=&\left(t_{001}-t_{0}t_{01}\right) +
\left(t_{011}-t_{1}t_{01}\right),  \cr
&&..., 
\label{z5}
\end{eqnarray}
the contribution from long orbits is mimicked by the combinations of the
short ones, which contribute to the sum with the opposite sign. As a result,
the magnitude of the coefficients $c_{i}$ of this expansion rapidly
decrease. This allows to estimate very effectively the asymptotics of the
coefficients of the expansion \cite[1]{a4}, and to prove the analyticity
properties stated above.

\section{Generalizations}

The basis of the PF operators theory was developed by Grothendieck in \cite
{a5}. His ideas were generalized recently by Ruelle \cite{a6} and Kitaev 
\cite{a7}, who proposed to consider formal sums or integrals of PF operators 
$L_{\omega}$, depending on a certain (discrete or continuous) parameter $%
\omega $: 
\begin{equation}
{\bf L}(x,y) {\bf =}\int \mu \left(d\omega \right) L_{\omega
}\left(x-f_{\omega}\left(y\right) \right) .  \label{z6}
\end{equation}
Here $\mu \left(d\omega \right) $ is some appropriate measure \cite{a6}.
One of the most important requirements imposed on such a sum or an integral,
is that the corresponding dynamical systems have to be fully chaotic for all
the values of $\omega \in \Omega $.

As shown in \cite{a6}, under certain natural requirements imposed on the
dynamical mapping function $f(x) $ and the amplitude, the
corresponding $\zeta $ functions and the Fredholm determinants are analytic
functions in a certain domain of the complex plane. However, the size of
this domain is smaller than the one in the case of a single PF operator.
Even if the individual systems $L_{\omega}\left(f_{\omega}\right) $ have
an entire Fredholm determinant $Z$ and the zeta function, the size of the
analyticity domain of the averaged operator $\left\langle L\right\rangle $
can be finite.

Physically, operators like (\ref{z5}) can be used to study chaotic systems
influenced by noise. Also, as shown recently in \cite{a8}, the quantum
mechanical Green's function of certain vector quantum particles can be
presented as a sum similar to (\ref{z6}). Recent series of publications
offered some effective methods for dealing with a system perturbed by
Gaussian noise, including that of smooth conjugations \cite{a9} and local
matrix representation \label{a10}\cite{a10}. However within these techniques
the connection with the cycle expansion is not transparent. In this paper it
is attempted to construct a cycle expansion technique for discrete versions
of such sums, for which an analogue with cycle expansion technique is easy
to establish.

\section{Double PF operator}

The simplest ``generalized PF\ operator'' is a formal sum of two
(noncommuting) PF operators (\ref{z2}), 
\begin{equation}
L^{(2)}(x,y) =A_{1}\delta \left(y_{\mu}-f_{1\mu
}^{t}(x) \right) +A_{2}\delta \left(y_{\mu}-f_{2\mu}^{t}(x) \right) ,  
\label{z7}
\end{equation}
Here $A_{1}$ and $A_{2}$ are certain multiplicative amplitudes.

Following the standard procedure \cite[2]{a1}, in order to describe the
asymptotics of the evolution determined by the operator $L^{(2)
}(x,y) $, we need to evaluate it's largest eigenvalue, or the
smallest root of the equation 
\begin{equation}
\det \left(1-zL^{(2)}\right) =\exp -\sum_{n=1}^{\infty}
\frac{z^{n}}{n}Tr\left(L^{(2)}\right) ^{n}=\exp \sum_{n=1}^{\infty}
\sum_{\left|i\right| =n}Tr\left(L_{i_{1}}....L_{i_{n}}\right) =0,
\end{equation}
where $i_{k}=1,2$. If a sequence $L_{i_{1}}....L_{i_{n}}$ ($i_{k}=1,2$) is
periodic, it can be written as a power of a shorter aperiodic string, 
\begin{equation}
L_{i_{1}}....L_{i_{n}}=\left(L_{i_{1}'}....L_{i_{s}'}\right) ^{q},
\end{equation}
where $n=s\cdot q$. Also, since only the trace of these operators is
considered, the sequences obtained from one another by cyclic permutations
give identical contributions, and therefore any length-$s$ string
contributes exactly $s$ times. Hence 
\begin{equation}
\sum_{\left| i\right| =n}Tr\left(L_{i_{1}}....L_{i_{n}}\right)
=\sum_{q=1}\sum_{\left| i\right| =s}s\cdot Tr\left(L_{i_{1}^{\prime
}}....L_{i_{s}'}\right) ^{q},\quad i_{1},...,i_{n}=1,2,
\end{equation}
and for the spectral determinant (\ref{z7}) one has: 
\begin{equation}
\log \det \left(1-zL^{(2)}\right) =\sum_{q=1}^{\infty}
\sum_{s=1}^{\infty}\sum_{\left| i\right| =s}\frac{z^{s\cdot q}}{q}Tr\left(
L_{i_{1}'}....L_{i_{s}'}\right) ^{q},
\end{equation}
where the sum is written in terms of the powers $q$ of the {\em aperiodic}
strings of the length $s$ of the symbols $L_{1}$ and $L_{2}$.

Using the formula 
\begin{equation}
\delta \left(y_{\mu}-f_{1,\mu}\left(z_{1}\right) \right) ...\delta
\left(z_{n,\mu}-f_{2,\mu}(x) \right) =\delta \left(y_{\mu
}-f_{1,\mu}\circ f_{2,\mu}\circ ...\circ f_{n,\mu}(x) \right)
, 
\end{equation}
one gets the following expression for the trace of a generic term in the
exponent: 
\begin{equation}
Tr\left(L_{i_{1}}...L_{i_{s}}\right) ^{q}=\int L_{i_{1}}\left(
x,y_{1}\right) ...L_{i_{n}}\left(y_{n-1},x\right) dxdy_{1}...dy_{n-1}=\int
\delta \left(x-f_{i_{1}}^{t}\left(y\right) \right) \times ... 
\end{equation}
\begin{equation}
...\times \delta \left(y-f_{i_{n}}^{t}(x) \right)
dxdy_{1}...dy_{n-1}=\int \delta \left(x-f_{i_{1}}\left(y\right) \circ
...\circ f_{i_{n}}(x) \right) =\sum_{w}\frac{1}{\left| 1-\Lambda
_{w,q}^{(s)}\right|}, 
\end{equation}
where $w$'s are the fixed points of the map $f_{i_{1}}\left(y_{1}\right)
\circ ...\circ f_{i_{n}}(x) $. The numbers $s_{1}$ and $s_{2}$
show how many times the operators $L_{1}$ and $L_{2}$ appear in the string 
$L_{i_{1}}....L_{i_{s}}$. Obviously $s_{1}+s_{2}=s$. Here for simplicity we
put $A_{1}=A_{2}=1$, to avoid unnecessary complications in formulas. The
dependence on the amplitudes is easy to reconstruct at the final step, so
they will be ignored until the very last section.

Hence, there is an infinite number of nonequivalent operators $%
L_{i_{1}'}....L_{i_{s}'}$ corresponding to each aperiodic
string, and we are looking at their fixed points. Proceeding as usual, we
can expand the denominator:

\begin{equation}
\frac{1}{\left| 1-\Lambda _{w,q}^{(s)}\right|}
=\sum_{w}\sum_{m=0}^{\infty}\frac{1}{\left| \Lambda _{w,q}^{(s)}\right| 
\Lambda _{w,q}^{(s) m}}. 
\end{equation}
Here $(s) $ denotes any of the aperiodic sequences of the length 
$s$, and $\Lambda _{p}^{(s)}$ represents the expansion factor
of a prime orbit indexed by ``$p$'' of a composite operator 
$L_{i_{1}}....L_{i_{s}}$. Substituting this expansion series into the
exponent yields the following Fredholm determinant: 
\begin{equation}
\det \left(1-zL\right) =\exp -\sum_{(s)}\sum_{p}\sum_{m=0}^{\infty}
\sum_{r=1}^{\infty}\frac{1}{r}
\left(\frac{z^{s\cdot n_{p}}}{\left| \Lambda _{p}^{(s)}\right| \Lambda
_{p}^{(s) m}}\right) ^{r}=\prod_{m=0}^{\infty}\prod_{(s)}\prod_{p}
\left(1-\frac{z^{s\cdot n_{p}}}{\left| \Lambda_{p}^{(s)}\right| 
\Lambda _{p}^{(s) m}}\right) . 
\end{equation}
Correspondingly, the product 
\begin{equation}
\zeta _{m}^{-1}=\prod_{(s)}\prod_{p}\left(1-\frac{z^{s\cdot
n_{p}}}{\left| \Lambda _{p}^{(s)}\right| 
\Lambda _{p}^{(s) m}}\right)  
\label{z8}
\end{equation}
can be considered as a generalization of the zeta function of the single-PF
operator (\ref{z4}).

This object is quite different from it's one-mapping counterpart. While the
single-mapping zeta function (\ref{z4}) is defined on the periodic orbits of
 the system (\ref{z1}), the product $\zeta _{m}^{-1}$ is taken over the fixed
points of the (aperiodic) combinations $f_{i_{1}'}\circ ...\circ
f_{i_{s}'}$, which can be thought of as ``interference'' terms.
Expanding the $(s) $ product in (\ref{z8}), one has 
\begin{equation}
\zeta _{m}^{-1}=\prod_{p}\left(1-\frac{z^{n_{p}}}{\left| 
\Lambda_{p}^{(1)}\right| \Lambda _{p}^{(1) m}}\right)
\left(1-\frac{\left(z^{2}\right) ^{n_{p}}}{\left| 
\Lambda _{p}^{(2)}\right| \Lambda _{p}^{(2) m}}\right) 
\left(1-\frac{\left(z^{3}\right) ^{n_{p}}}{\left| 
\Lambda _{p}^{(3)}\right|\Lambda _{p}^{(3) m}}\right) ...\ .  
\label{z9}
\end{equation}

In order to avoid unnecessary complication of formulas, the repetition index 
$m$ will be suppressed in the sequel. The parentheses $(1)$, $(2)$, $(3)$ and 
so on imply the product over all the possible aperiodic combinations 
$L_{i_{1}}....L_{i_{s}}$, $i=1,2$, of
lengths $1$, $2$, $3$ correspondingly. For instance, the first bracket
consists of the simple product of two $\zeta $-functions (two possible
length-$1$ strings), 
\begin{equation}
\prod_{p}\left(1-z^{n_{p}}t_{p}^{(1)}\right) =\prod_{p}
\left(1-z^{n_{p}}t_{p}^{1}\right) \prod_{p}\left(1-z^{n_{p}}t_{p}^{2}\right) , 
\end{equation}
where $t_{p}^{i}\equiv \frac{1}{\left| \Lambda _{p}^{i}\right| \Lambda
_{p}^{m}}$. This is just the product of the individual $\zeta $-functions, 
\begin{eqnarray*}
\zeta ^{-1}\left(z\mid L_{1}\right) &=&\prod_{p}
\left(1-z^{n_{p}}t_{p}^{1}\right) =1-t_{f}^{1}z-c_{2}^{1}z^{2}-c_{3}^{1}z^{3}...,
\cr
\zeta ^{-1}\left(z\mid L_{2}\right) &=&\prod_{p}
\left(1-z^{n_{p}}t_{p}^{2}\right) =1-t_{f}^{2}z-c_{2}^{2}z^{2}-c_{3}^{2}z^{3}...\ .
\end{eqnarray*}
The term $t_{f}^{1}$ and $t_{f}^{1}$ represent the fundamental cycles, and
the coefficients $c_{i}^{1}$ and $c_{i}^{2}$ denote the curvature
corrections for the first and the second mapping correspondingly.

The only length-two string $L_{1}L_{2}$ (the first ``interference''
correction) produces it's own zeta function, which contributes the term 
\begin{equation}
\zeta ^{-1}\left(z\mid L_{1}L_{2}\right) 
=\prod_{p}\left(1-z^{2n_{p}}t_{p}^{12}\right)
=1-t_{f}^{12}z^{2}-c_{2}^{12}z^{4}-c_{3}^{12}z^{6}... 
\end{equation}
to the product. Here $t_{f}^{12}$, $c_{2}^{12}$, $c_{3}^{12}$, etc., are the
fundamental cycles and the curvature corrections for the mapping 
$f_{1}\circ f_{2}(x) $.

The two length-three strings $L_{1}L_{1}L_{2}$ and $L_{1}L_{2}L_{2}$
contribute
\begin{equation}
\zeta ^{-1}\left(z\mid L_{1}L_{1}L_{2}\right) 
=\prod_{p}\left(1-z^{3n_{p}}t_{p}^{112}\right)
=1-z^{3}t_{f}^{112}-c_{2}^{112}z^{6}-c_{3}^{112}z^{9}..., 
\end{equation}
and 
\begin{equation}
\zeta ^{-1}\left(z\mid L_{1}L_{2}L_{2}\right) 
=\prod_{p}\left(1-z^{n_{p}}t_{p}^{122}\right)
=1-z^{3}t_{f}^{122}-c_{2}^{122}z^{6}-c_{3}^{122}z^{9}\ ..., 
\end{equation}
correspondingly, and so on. The top index shows the aperiodic strings of
operators $L_{1}$ and $L_{2}$, while the lower index enumerates the prime
orbits of each operator. $t_{f}^{(s)}$ represents the
fundamental cycle of a given $(s) $ operator, 
$L_{i_{1}}....L_{i_{s}}$, and the coefficients $c_{n}^{(s)}$
give the corresponding curvature corrections.

It is important to notice, that the symbolic dynamics of the individual
mappings, used to obtain the expressions for $\zeta ^{-1}\left(z\mid L_{i}\right)$, 
$\zeta^{-1}\left(z\mid L_{1}L_{2}\right)$, 
$\zeta^{-1}\left(z\mid L_{1}L_{1}L_{2}\right)$, 
$\zeta ^{-1}\left(z\mid L_{1}L_{2}L_{2}\right)$, etc., do not need to be specified. 
In fact, one might use different symbolic dynamics in order to obtain the power 
expansion for these functions. After the necessary expansion series are at hand, 
the details of the symbolic dynamics of the corresponding mappings are
irrelevant. The further analysis is made entirely in terms of the 
$t_{f}^{(s)}$s and $c_{n}^{(s)}$s.

Let us consider the power expansion of all the zeta-products up to the third
power of $z$. The expansion for the two-PF operator will be then: 
\begin{eqnarray}
\zeta _{2}^{-1}(z) =
\left(1-t_{f}^{1}z-c_{2}^{1}z^{2}-c_{3}^{1}z^{3}-O\left(z^{4}\right)\right)
\left(1-t_{f}^{2}z-c_{2}^{2}z^{2}-c_{3}^{2}z^{3}-O\left(z^{4}\right)\right) 
\cr\times 
\left(1-z^{2}t_{f}^{12}-O\left(z^{4}\right) \right) 
\left(1-z^{3}t_{f}^{112}-O\left(z^{6}\right)\right) 
\left(1-z^{3}t_{f}^{122}-O\left(z^{6}\right)\right)
\cr
=1-\left(t_{f}^{1}+t_{f}^{2}\right) z-\left(\left(
t_{f}^{12}-t_{f}^{1}t_{f}^{2}\right) +c_{2}^{1}+c_{2}^{2}\right) z^{2}
\cr
-\Big[\left(t_{f}^{122}-t_{f}^{2}t_{f}^{12}\right) 
-\left(t_{f}^{112}-t_{f}^{1}t_{f}^{12}\right) 
-\left(t_{f}^{1}+t_{f}^{2}\right)
\left(c_{2}^{1}+c_{2}^{2}\right)
\cr
+t_{f}^{1}c_{2}^{1}+t_{f}^{2}c_{2}^{2}+c_{3}^{1}+c_{3}^{2}\Big]z^{3}+
O\left(z^{4}\right),
\label{z10} 
\end{eqnarray}
and so on, where $t_{f}^{i}$'s and $c_{n}^{i}$ are the fundamental
contributions and the curvature corrections of the individual mappings.

Comparing the expressions (\ref{z10}) and (\ref{z4},\ref{z5}), one can 
see that the structure of the coefficients $t_{f}$ and $c_{n}$ is 
repeated on a ``new level''. 
The new ``fundamental'' contribution is now 
\begin{equation}
T_{f}^{(2)}=t_{f}^{1}+t_{f}^{2}, 
\end{equation}
and the following terms have a structure similar to the corresponding
``curvature corrections'' in the case of a system with complete binary
dynamics (\ref{z5}). For example the term 
\begin{equation}
C_{2}\equiv t_{f}^{12}-t_{f}^{1}t_{f}^{1}, 
\end{equation}
is formally identical to the $c_{2}$ of (\ref{z5}), the term $C_{3}$ also
contains terms 
\begin{equation}
\left(t_{f}^{122}-t_{f}^{2}t_{f}^{12}\right)+
\left(t_{f}^{1}t_{f}^{12}-t_{f}^{112}\right) 
\end{equation}
of manifestly ``curvature correction'' type.

However, starting from $C_{3}$ there start to appear some additional
``curvature combinations'', such as 
\begin{equation}
\left(t_{f}^{1}+t_{f}^{2}\right) \left(c_{2}^{1}+c_{2}^{2}\right)
-t_{f}^{1}c_{2}^{1}-t_{f}^{2}c_{2}^{2}. 
\end{equation}
Also, every term contains the sum of all the standard curvature coefficients 
$c_{n}$ of the corresponding degree, associated with the ``pure'' maps $%
f_{1}(x) $ and $f_{2}(x) $.

In short notations, which are going to be used below, the expansion 
(\ref{z10}) can be written as: 
\begin{equation}
\zeta _{2}^{-1}(z) =1-zT_{f}-z^{2}\left(
C_{2}+\sum_{i=1}^{2}c_{2}^{i}\right) -z^{3}\left(
C_{3}-\sum_{i=1}^{2}t_{f}^{i}\sum_{i=1}^{2}c_{2}^{i}+
\sum_{i=1}^{2}t_{f}^{i}c_{2}^{i}+\sum_{i=1}^{2}c_{3}^{i}\right) 
+O\left(z^{4}\right).
\label{z11}
\end{equation}
It should be emphasized, that the coefficients $T_{f}$ and $C_{n}$ are
principally different from their usual, single PF counterparts $t_{f}$ and 
$c_{n}$. They are constructed using the contributions of {\em different}
mappings, (not only the $f_{1}(x) $ and $f_{2}(x) $
but also from all their aperiodic superpositions), and therefore produce
some ``interference'' effect.

\section{Triple PF operator}

One can easily write down the explicit form of the power expansion for the
triple PF operator 
\begin{equation}
L^{(2)}(x,y) =A_{1}\delta \left(y_{\mu}
-f_{1\mu}^{t}(x) \right) +A_{2}\delta \left(y_{\mu}
-f_{2\mu}^{t}(x) \right) +A_{3}\delta \left(y_{\mu}
-f_{3\mu}^{t}(x) \right) .  
\label{z12}
\end{equation}
As in the double PF operator case, the expansion is made up to the third
order:
\begin{eqnarray}
\zeta _{3}^{-1}=\prod_{i=1}^{3}
\left(1-t_{f}^{i}z-c_{2}^{i}z^{2}-c_{3}^{i}z^{3}-O(z^{4}) \right)
\prod_{ij}^{3}\left(1-z^{2}t_{f}^{ij}-O(z^{4})\right)
\prod_{ijk}^{3}\left(1-z^{3}t_{f}^{ijk}-O(z^{6}\right) \cr
=1-z\left(t_{f}^{1}+t_{f}^{2}+t_{f}^{3}\right) 
-z^{2}\left(
 \left(t_{f}^{23}-t_{f}^{2}t_{f}^{3}\right) 
+\left(t_{f}^{13}-t_{f}^{1}t_{f}^{3}\right) 
+\left(t_{f}^{12}-t_{f}^{1}t_{f}^{2}\right) 
+c_{2}^{1}+c_{2}^{2}+c_{2}^{3}\right) 
\nonumber
\end{eqnarray}
\begin{eqnarray}
+z^{3}\Bigg[\left(t_{f}^{223}-t_{f}^{2}t_{f}^{23}\right) 
+\left(t_{f}^{233}-t_{f}^{3}t_{f}^{23}\right) 
+\left(t_{f}^{112}-t_{f}^{1}t_{f}^{12}\right) 
+\left(t_{f}^{122}-t_{f}^{2}t_{f}^{12}\right) 
\cr
+\left(t_{f}^{113}-t_{f}^{1}t_{f}^{13}\right) 
+\left(t_{f}^{133}-t_{f}^{3}t_{f}^{13}\right) 
+t_{f}^{213}+t_{f}^{132}-t_{f}^{3}t_{f}^{12}-t_{f}^{2}t_{f}^{13}
-t_{f}^{1}t_{f}^{23}+t_{f}^{123}+t_{f}^{1}t_{f}^{2}t_{f}^{3}
\cr
-t_{f}^{1}\left(c_{2}^{2}+c_{2}^{3}\right) 
-t_{f}^{2}\left(c_{2}^{1}+c_{2}^{3}\right)
-t_{f}^{3}\left(c_{2}^{1}+c_{2}^{3}\right) +
\left(c_{3}^{1}+c_{3}^{2}+c_{3}^{3}\right)\Bigg]+O(z^{4}) , 
\end{eqnarray}
where $t_{f}$, $c_{n}$ are the fundamental contributions and the curvature
corrections of the corresponding mappings.

It is easy to observe that the above expansion is very similar to the one
obtained for a system with a complete ternary symbolic dynamics, for which
the expansion coefficients are:
\begin{eqnarray*}
t_{f}^{(3)} &=&t_{3}+t_{2}+t_{1} \cr c_{2}^{(3)} 
&=&
\left(t_{13}-t_{1}t_{3}\right) 
+\left(t_{23}-t_{2}t_{3}\right) 
+\left(t_{12}-t_{1}t_{2}\right) \cr
c_{3}^{(3)} &=&
\left(t_{233}-t_{3}t_{23}\right) 
+\left(t_{223}-t_{2}t_{23}\right) 
+\left(t_{133}-t_{3}t_{13}\right) 
+\left(t_{113}-t_{1}t_{13}\right) 
+\left(t_{122}-t_{2}t_{12}\right) \cr
&&
+\left(t_{112}-t_{1}t_{12}\right)
+t_{213}+t_{123}-t_{3}t_{12}-t_{2}t_{13}-\allowbreak
t_{1}t_{23}+t_{1}t_{2}t_{3}
\end{eqnarray*}

Using the notations $T_{f}^{(3)}$, $C_{2}^{(3)}$, $C_{3}^{(3)}$ for the 
corresponding combinations of prime contributions in (\ref{z14}), we get 
for the triple operator zeta functions: 
\begin{equation}
\zeta _{3}^{-1} =1-zT_{f}^{(3)}
-z^{2}\left(C_{2}^{(3)}+\sum_{i=1}^{3}c_{2}^{i}\right) 
-z^{3}\left(C_{3}^{(3)}
-\sum_{i=1}^{3}t_{f}^{i}\sum_{i=1}^{3}c_{2}^{i}
+\sum_{i=1}^{3}t_{f}^{i}c_{2}^{i}
+\sum_{i=1}^{3}c_{3}^{i}\right) +O(z^{4}). 
\end{equation}
Comparing this to the expansion of the double PF operator, 
\begin{equation}
\zeta _{2}^{-1}=1-zT_{f}^{(2)}
-z^{2}\left(C_{2}^{(2)}+\sum_{i=1}^{2}c_{2}^{i}\right) 
-z^{3}\left(C_{3}^{(2)}-\sum_{i=1}^{2}t_{f}^{i}
\sum_{i=1}^{2}c_{2}^{i}+
\sum_{i=1}^{2}t_{f}^{i}c_{2}^{i}+\sum_{i=1}^{2}c_{3}^{i}\right) 
+O(z^{4}) , 
\end{equation}
it is easy to observe that the curvature coefficients have the same
structure, depending trivially on the number of the PF operators in the sum
elements. The main difference is contained in the terms $C_{2}^{(3)}$, 
$C_{3}^{(3)}$ as opposed to $C_{2}^{(2)}$, $C_{3}^{(2)}$, because they are 
constructed as the curvature terms of systems with binary and ternary symbolic 
dynamics correspondingly.

\section{$N$-sums and the continuous limit}

Previous analysis can be conducted in the exact same way for any $N$
-operator sum ($N\geq 2$), 
\begin{equation}
L^{(N)}(x,y) =\sum_{i=1}^{N}\delta \left(y_{\mu}-f_{i,\mu}^{t}(x)\right) .  
\label{z14}
\end{equation}
For the Fredholm determinant $\det \left(1-zL\right) $ one has: 
\begin{equation}
\det \left(1-zL\right) =\exp \left(-\sum_{n=1}^{\infty}\frac{z^{n}}{n}
Tr\left(L^{n}\right) \right) =\prod_{m}\prod_{s}\prod_{p}
\left(1-\frac{z^{s\cdot n_{p}}}{\left| \Lambda _{p}^{(s)}\right| 
\Lambda_{p}^{(s) m}}\right).
\label{z15}
\end{equation}

Based on the previous examples and using the induction method, one can prove
that the power expansion of the zeta function of an $N$-term PF operator is 
\begin{eqnarray}
\zeta _{N}^{-1}(z) =1-zT_{f}^{(N)}
-z^{2}\left(C_{2}^{(N)}+\sum_{i=1}^{N}c_{2}^{i}\right)
\cr
-z^{3}\left(C_{3}^{(N)}+\sum_{i=1}^{N}t_{f}^{i}\sum_{i=1}^{N}c_{2}^{i}
-\sum_{i=1}^{N}t_{f}^{i}c_{2}^{i}+\sum_{i=1}^{N}c_{3}^{i}\right) 
+O(z^{4}),
\label{z16}
\end{eqnarray}
where 
\begin{equation}
T_{f}^{(N)}=\sum_{i=1}^{N}t_{f}^{i}, 
\end{equation}
and $C_{2}^{(N)}$, $C_{3}^{(N)}$, etc., have the
structure identical to the curvature correction coefficients of a system
with a complete $N$-ary symbolic dynamics. The extra terms which appear in
the expansion (\ref{z16}), 
\begin{eqnarray}
\tilde{c}_{2}^{(N)} &\equiv &\sum_{i=1}^{N}c_{2}^{i},
\label{z17} \cr
\tilde{c}_{3}^{(N)} &\equiv&
\sum_{i=1}^{N}t_{f}^{i}\sum_{i=1}^{N}c_{2}^{i}-
\sum_{i=1}^{N}t_{f}^{i}c_{2}^{i}+\sum_{i=1}^{N}c_{3}^{i},
\end{eqnarray}
\begin{equation}
..., 
\end{equation}
can be called ``direct curvature contributions'' as opposed to the
``interference curvatures'' $C_{n}^{(N)}$.

The direct curvature terms (\ref{z17}) depend trivially on the number of PF
operators included into the sum (\ref{z15}). Among the interference
curvature terms, only the fundamental contribution term $T_{f}$ is just the
direct sum of the individual fundamental contributions.

It is important to emphasize, that the complete curvature coefficients have
a {\em universal form}, 
\begin{equation}
\tilde{C}_{n}^{(N)}=C_{n}^{(N)}+\tilde{c}_{n}^{(N)},  
\label{z18}
\end{equation}
which provides an algorithm for evaluating the zeta-function expansion
coefficients of the multiple PF operators to a given degree $n$. Moreover,
the direct curvature coefficients depend trivially on the order of the sum 
(\ref{z15}). This allows to make certain statements about the behavior of the 
{\em continuous} sums, 
\begin{equation}
{\bf L}(x,y) =\int \mu \left(d\omega \right) \ A_{\omega}
\delta \left(y_{\mu}-f_{\mu}^{t}\left(x,\omega \right) \right) ,
\label{z19}
\end{equation}
from the point of view of the standard cycle expansion technique. Here the
amplitudes $A_{\omega}$ and the mapping functions $f_{\mu}^{t}
\left(x,\omega \right) $ depend continuously on the parameter $\omega $.

After some elementary considerations it is easy to reconstruct the explicit
amplitude dependence in the expansions of the MPF operators' zeta functions, 
\begin{equation}
\zeta _{N}^{-1}=\prod_{(s)}\prod_{p}\left(1-\frac{z^{s\cdot n_{p}}
A^{(s) n_{p}}}{\left| \Lambda _{p}^{(s)}\right| 
\Lambda _{p}^{(s) m}}\right) ,  
\label{z20}
\end{equation}
where $A^{(s)}\equiv A_{1}^{s_{1}}...A_{2}^{s_{N}}$. The
product over the $(s) $ includes all the possible strings in
which the symbols $L_{1}$,..., $L_{N}$ appear $s_{1}$,..., $s_{2}$ times
correspondingly, for all $s_{1}+...+s_{N}=s$. The expansion of (\ref{z20})
yields: 
\begin{eqnarray*}
\zeta _{N}^{-1} &=&1-z\sum_{i=1}^{N}A_{i}t_{f}^{i}-z^{2}\left(A^{\left(
2\right)}C_{2}+\sum_{i=1}^{N}A_{i}^{2}c_{2}^{i}\right) \cr
&&-z^{3}\left(A^{(3)
}C_{3}+\sum_{i=1}^{N}A_{i}^{3}c_{3}^{i}-\sum_{i=1}^{N}A_{i}t_{f}^{i}%
\sum_{i=1}^{N}A_{i}^{2}c_{2}^{i}+\sum_{i=1}^{N}A_{i}^{3}t_{f}^{i}c_{2}^{i}%
\right) +O(z^{4}) .
\end{eqnarray*}

Without loss of generality, one can assume that there exists such a set of
discrete parameters $\omega _{1},...,\omega _{N}$, that the curvature
corrections of the individual maps can be written as: 
\begin{eqnarray}
t_{f}^{i} &\equiv &t_{f}\left(\omega _{i}\right) ,  \cr
c_{n}^{i} &\equiv &c_{n}\left(\omega _{i}\right) .
\label{a21}
\end{eqnarray}
In case if the amplitudes $A_{\omega}$ and the mapping functions $f_{\mu
}^{t}\left(x,\omega \right) $ in (\ref{z20}) depend continuously on the
parameter $\omega $, the coefficients (\ref{z19}) also depend continuously
on $\omega $, at least for a certain range of parameters. Therefore, the
fundamental contribution coefficient will produce in the limit $N\rightarrow
\infty $: 
\begin{equation}
T_{f}^{(N)}=\sum_{i=1}^{N}A_{i}t_{f}^{i}\rightarrow T_{f}=\int
t_{f}(\omega) \mu \left(d\omega \right) . 
\end{equation}
The direct curvature contributions in the expansion (\ref{z17}) also can be
written in the integral form: 
\begin{eqnarray*}
\tilde{c}_{2}^{(N)} &\equiv
&\sum_{i=1}^{N}A_{i}^{2}c_{2}^{i}\equiv \sum_{i=1}^{N}A_{i}^{2}c_{2}
\left(\omega _{i}\right) \rightarrow \tilde{c}_{2}\equiv 
\int c_{2}\left(\omega\right) \mu \left(d^{2}\omega \right) , \cr
\tilde{c}_{3} &\equiv &\int t_{f}(\omega) \mu \left(d\omega
\right) \int c_{2}(\omega) \mu \left(d^{2}\omega \right) -\int
t_{f}c_{2}\mu \left(d^{3}\omega \right) +\int c_{3}(\omega)
\mu \left(d^{3}\omega \right) ,
\end{eqnarray*}
\begin{equation}
...\ . 
\end{equation}

On the other hand, the interference curvature coefficients $C_{n}^{\left(
N\right)}$ the limit $N\rightarrow \infty $ are considerably more
complicated, since the elements contained in the higher order interference
curvatures $C_{n}^{(N)}$, $n\geq 3$, are labeled by the
aperiodic strings {\em of complexity}$N$.

However, the $N\rightarrow \infty $ limit for the interference coefficients
exist \cite{a9,a10}, and therefore zeta function $\zeta ^{-1}$ has the
following expansion form: 
\begin{eqnarray}
\zeta _{N\rightarrow \infty}^{-1} =1-z\int t_{f}(\omega) \mu
\left(d\omega \right) -z^{2}\left(C_{2}^{N}+\int c_{2}
\left(\omega\right) \mu \left(d^{2}\omega \right) \right) \cr
-z^{3}\Bigg(C_{3}^{N}+\int t_{f}(\omega) \mu \left(d\omega
\right) \int c_{2}(\omega) \mu \left(d^{2}\omega \right) +\int
t_{f}c_{2}(\omega) \mu \left(d^{3}\omega \right) 
\cr
-\int c_{3}(\omega) \mu \left(d^{3}\omega \right) \Bigg)
+O(z^{4}) .
\end{eqnarray}
This expansion is analogous to the cumulant expansion of 
$\det \left(1-zL\right) $ obtained in \cite{a10}.

\section{Conclusion}

From practical as well as from theoretical point of view, it is certainly
more convenient to treat the continuous sum case using the methods developed
in \cite{a9,a10}, because these methods effectively go around some convoluted
calculations, which are necessary to evaluate the expansion coefficients (%
\ref{z17}). However, for discrete versions of such ``noisy'' PF operators,
the multiple sums (\ref{z14}), these methods can not be applied directly.
Meanwhile, as it was shown above, such systems allow a direct treatment
which is analogous to the standard cycle expansion technique. The detailed
form of the corresponding expansion coefficients (\ref{z17}) reveals a
remarkable structural simplicity and shows certain a familiar features
characteristic for the single-operator expansion coefficients.

In practice, even in case of the a single PF operator, one can usually
obtain only the first several terms of the cycle expansion series. In case
of multiple PF operators, the explicit form of the generalized curvatures (%
\ref{z17}) conveniently illustrates the details of the interference between
the different maps constituting a multiple PF operator, using the language
of the periodic cycles arising from the maps themselves as well as their
compositions. This provides a nontrivial extension of the periodic orbit
theory to a more complicated dynamical evolution which involves a few
''evolution channels''.

\section{Acknowledgments}

I am grateful to Professor R. Artuso for reading the manuscript and making
useful comments.

The author gratefully acknowledges financial support by NSF
grants PHY-9900730 and PHY-9900746.

\end{document}